\documentclass{article}

\usepackage{arxiv}

\usepackage[utf8]{inputenc} % allow utf-8 input
\usepackage[T1]{fontenc}    % use 8-bit T1 fonts
\usepackage{hyperref}       % hyperlinks
\usepackage{url}            % simple URL typesetting
\usepackage{booktabs}       % professional-quality tables
\usepackage{amsfonts}       % blackboard math symbols
\usepackage{nicefrac}       % compact symbols for 1/2, etc.
\usepackage{microtype}      % microtypography
\usepackage{lipsum}
\usepackage{graphicx}
\graphicspath{ {./images/} }
\usepackage{blindtext}
\usepackage[square,sort]{natbib}

\title{Hardness Test of GRB 950830 as a Gravitationally Lensed Echo}

\author{
 Oindabi Mukherjee \\
  Michigan Technological University\\
  1400 Townsend Dr\\
  Houghton, MI 49931\\
  \texttt{omukherj@mtu.edu} \\
   \And
 Robert J. Nemiroff \\
  Michigan Technological University\\
  1400 Townsend Dr\\
  Houghton, MI 49931\\
  \texttt{nemiroff@mtu.edu} \\
  \And
}

\begin{document}
\maketitle
\begin{abstract}
Cumulative hardness comparisons are a simple but statistically powerful test for the presence of gravitational lensing in gamma-ray bursts (GRBs). Since gravitational lensing does not change photon energies, all source images should have the same spectra -- and hence hardness. Applied to the recent claim that the two pulses in GRB 950830 are lensed images of the same pulse, the measured flux ratio between the two main pulses should be the same at all energies. After summing up all the counts in both of GRB 950830's two pulses in all four BATSE energy bands, it was found that in energy channel 3, the second pulse appears somewhat weak. In comparison with the other BATSE energy channels, the difference was statistically significant at above 90\%. This model-independent test indicates that the case for GRB 950830 involving a gravitational lens may be intriguing -- but should not be considered proven.
\end{abstract}

% keywords can be removed

\section{Introduction} \label{sec:intro}
Recently a paper by 
\citet{paynter2021evidence} (PWT) claimed to identify a possible signature of a gravitational lens in the light curve of GRB 950830. PWT claimed that the first pulse near trigger time may be gravitationally echoed by a second pulse that starts shortly thereafter, thereby indicating that both pulses are gravitationally-lensed images of the same single source pulse. PWT's search for millilensing followed previous efforts starting with  \citet{Marani1999GrbMillilensing} and \citet{Nemiroff2001Millilensing}. PWT's Bayesian analysis indicates that a model that includes a gravitational lens fits the data better than a model that does not. 

In this work, we perform an additional test of the gravitational lens hypothesis of GRB 950830 (BATSE trigger 3770). We make use of the conjecture that the gravitational deflection of a photon does not change that photon's energy so that the gravitational lensing magnification of a source should be the same at every wavelength \citep{Paczynski1986GrbLensing}. However, instead of fitting the light curves to pulse forms, we create a model-independent sum of the total counts for each pulse in all four energy channels. We then test to see if the ratio of the counts in the two pulses is the same across all energy channels.

\section{Analysis} \label{sec:analysis}
Data from the Burst and Transient Source Experiment (BATSE) onboard the Compton Gamma-Ray Observatory was analyzed, specifically the same data set used by PWT. BATSE measured counts, typically an integer number of incident photons, which together were proportional to the measured incident flux. Throughout this analysis and following PWT, we used 5-millisecond bins. The four BATSE energy channels utilized were channel 1 (20 - 60 keV), channel 2 (60 - 110 keV), channel 3 (110 - 320 keV) and channel 4 (320 - 2000 keV).

To estimate the cumulative hardness comparisons, first, the background level was fit. It was assumed that the background level was constant during the one-second duration of this entire short GRB. The background levels determined were $B = 833, 658, 632, and, 439$ for energy channels 1, 2, 3, and 4 respectively. It was found that our final results were not sensitive to these background levels to an error of about 100 counts.

We next determined the starting and ending times of the first pulse from the summed counts across all energies. The starting time was chosen to be the time when these summed counts in 5-ms bins increased to over 1-$\sigma$ above the background fit. The pulse was considered to continue until the summed counts dropped to below 1-$\sigma$. It was assumed that each of the two pulses started at the same time in all energy bands \citep{Nemiroff2000PulseStartConjecture, Hakkila2009GrbPulseStart}.
The starting time of the second pulse was found by computing the off-set time that minimized $\chi^2$. The resulting starting times of the two pulses were determined to be -0.0634 and 0.3466 seconds relative to the BATSE trigger time, respectively. It was found that the results are not sensitive to the pulse start times within about 0.0048 seconds. 

This method also gave the duration of the pulses. Although each of the two pulses in each of the four channels might have different durations, for simplicity and to be consistent with a gravitational lens interpretation, both pulses were taken to have the same duration as that determined for the first pulse across all energies: 0.175 seconds. It was found that our final results were not sensitive to this duration to within about 0.0048 seconds.

Given these background fits, pulse start times, and pulse durations, the counts $P$ above background in both pulses and in all energy bands were calculated. This method allows pulse counts to be calculated in a pulse model-independent manner. The results are $P_{1,1} = 262$, $P_{2,1} = 243$, $P_{1,2} = 622$, $P_{2,2} = 496$, $P_{1,3} = 1303$, $P_{2,3} = 844$, $P_{1,4} = 220$, and, $P_{2,4} = 221$, where the first subscript refers to the pulse number and the second subscript refers to the energy channel.

\begin{figure}
    \includegraphics[width=0.9 \columnwidth, angle=0]{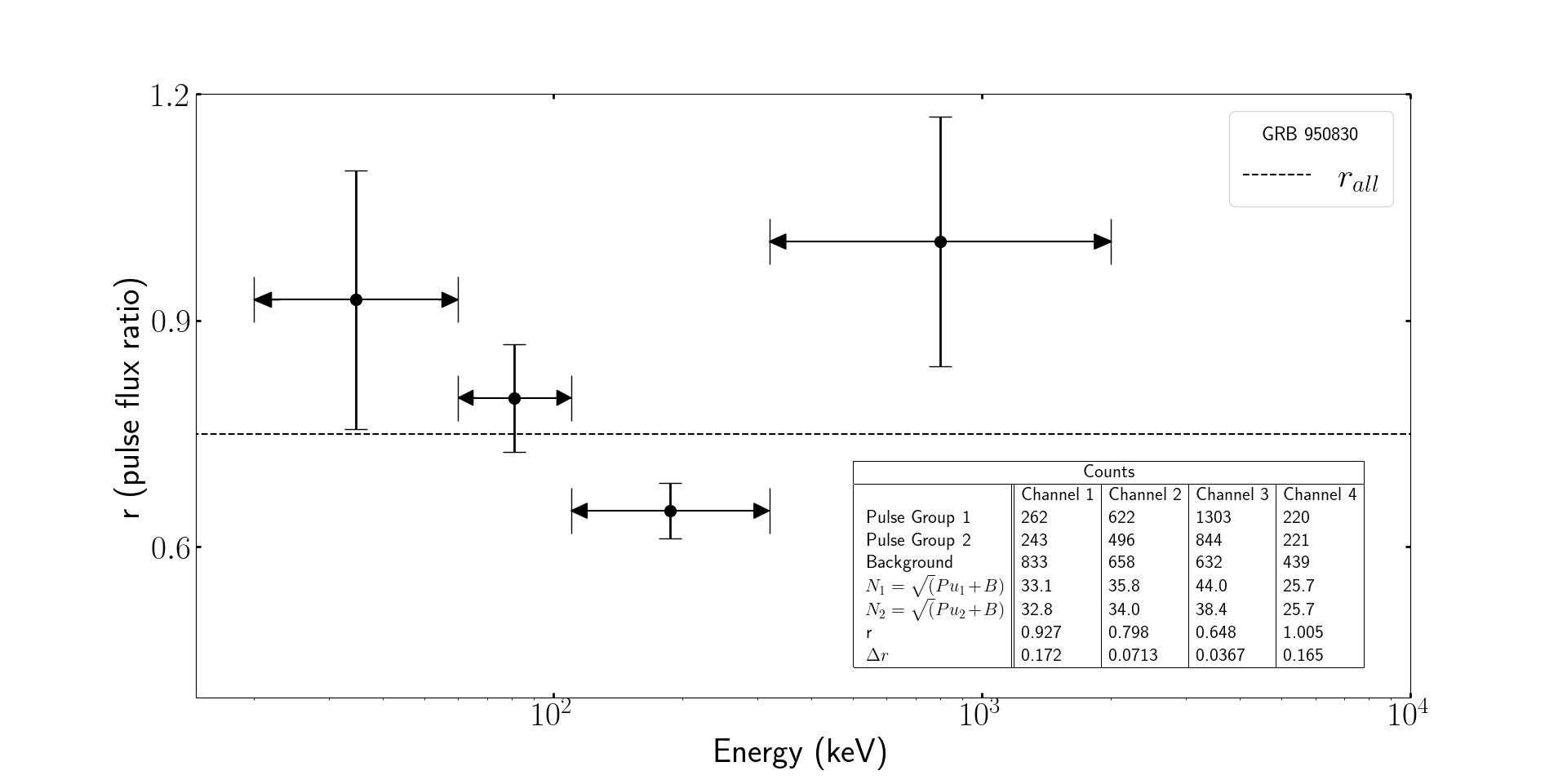}
    \caption{Count ratios between the two main pulses as a function of energy channel for GRB 950830}
    \label{fig:Figure_1}
\end{figure}

The ratio $r$ between the counts for the two pulses in each of the four energy channels was then found to be $r_1 = P_{2,1} / P_{1,1} = 0.927 \pm 0.172$ for channel 1, $r_2 = 0.798 \pm 0.0713 $ for channel 2, $r_3 = 0.648 \pm 0.0367 $ for energy channel 3, and, $r_4 = 1.005 \pm 0.165 $ for energy channel 4. The errors on these ratios were based on the Poisson noise inherent in the backgrounds as well as the pulses \citep{Sampling1977Cochran}. These $r$ values are shown in the table on the right side of Figure~\ref{fig:Figure_1}.

As seen in Figure~\ref{fig:Figure_1}, the $r$ value for channel 3 is somewhat lower than the other energy channels, whereas gravitational lensing demands that they be the same. Formally, the ratio ($r$) between counts of the two pulses in BATSE energy channel 3 differs from the same ratio in channel 4 at about the 2.11-$\sigma$ level (96.52 percent). Furthermore, the ratio between counts of the two pulses in BATSE energy channel 2 differs from the same pulse ratio in channel 3 at the 1.87-$\sigma$ level (93.86 percent). Additionally, summing counts across all energies and computing the mean $r_{all}$, it was found that the summed $\chi^2$ difference of the $r$ values from $r_{all}$ was 7.68, indicating a 94.7 \% probability of a significant difference. 

These cumulative hardness discrepancies appear to indicate that the case for GRB 950830 involving a gravitational lens may well be considered intriguing -- but should not be considered proven.

We thank Michigan Technological University for their general support, and Jon Hakkila and Neerav Kaushal for comments.

\bibliographystyle{unsrtnat}  
\bibliography{GrbNote}

\end{document}